\documentclass[conference]{IEEEtran}
\usepackage{cite}
\usepackage{amsmath,amssymb,amsfonts}
\usepackage{algorithmic}
\usepackage{graphicx}
\usepackage{textcomp}
\usepackage{xcolor}

\def\BibTeX{{\rm B\kern-.05em{\sc i\kern-.025em b}\kern-.08em
    T\kern-.1667em\lower.7ex\hbox{E}\kern-.125emX}}

\usepackage{multirow} 
\usepackage[hyphens]{url}

\usepackage[]{todonotes}

\begin{document}
	
\noindent
© 2021 IEEE.  Personal use of this material is permitted.  Permission from IEEE must be obtained for all other uses, in any current or future media, including reprinting/republishing this material for advertising or promotional purposes, creating new collective works, for resale or redistribution to servers or lists, or reuse of any copyrighted component of this work in other works.

\title{Towards Cross-Blockchain Smart Contracts
}

\author{\IEEEauthorblockN{Markus Nissl, Emanuel Sallinger}
	\IEEEauthorblockA{\textit{Database and Artificial Intelligence Group} \\
		\textit{TU Wien}, Vienna, Austria \\
		\{mnissl, sallinger\}@dbai.tuwien.ac.at}
	\and
	\IEEEauthorblockN{Stefan Schulte}
	\IEEEauthorblockA{\textit{Christian-Doppler-Laboratory}\\
		\textit{for Blockchain Technologies}\\
		\textit{for the Internet of Things} \\
		\textit{TU Wien}, Vienna, Austria \\
		s.schulte@dsg.tuwien.ac.at}
	\and
	\IEEEauthorblockN{Michael Borkowski}
	\IEEEauthorblockA{\textit{Institute of Flight Guidance} \\
		\textit{German Aerospace Center (DLR)}\\
		Brunswick, Germany \\
		michael.borkowski@dlr.de}
}

\maketitle

\begin{abstract}
In recent years, manifold blockchain protocols have been proposed by researchers and industrial companies alike. This has led to a very heterogeneous blockchain landscape. Accordingly, it would be desirable if blockchains could interact with each other. However, current protocols offer only limited support for interoperability, thus preventing tokens or smart contracts from leaving the scope of a particular blockchain. 

As a first step towards cross-chain smart contract interactions, we introduce a framework which allows to invoke a smart contract from another blockchain. 
We offer support for continuing a smart contract after receiving a result from a different blockchain, and for calling smart contracts recursively across blockchains. 
We provide a reference implementation for Ethereum-based blockchains using Solidity and evaluate the performance regarding time and cost overheads.
\end{abstract}


\section{Introduction}

\noindent
Since the advent of Bitcoin~\cite{Nakamoto_bitcoin}, blockchains have received significant attention by the industry and by research groups~\cite{Zohar:2015:BUH:2817191.2701411}. The manifold activities within the community have led to substantial progress in the blockchain realm: While Bitcoin only has limited features, novel blockchain implementations which provide more sophisticated functionalities arise continually.
These technological advances improve the cryptocurrency itself, e.g.,~by adding multi-signature support to Bitcoin, or by creating forks, 
e.g.,~Litecoin\cite{Litecoin}, use cryptocurrency technology as a protocol layer and add new layers on top of it, like
the Lightning Network~\cite{lightning}, or create completely new cryptocurrencies with novel concepts, such as smart contracts, which are well known, e.g.,~from Ethereum~\cite{ethereum_whitepaper, DBLP:journals/access/BorkowskiSFHS19}.

Hence, today's blockchain landscape is very fragment\-ed~\cite{zheng18}. This makes it very difficult for developers and researchers to track the progress of blockchains and to choose the most promising technologies. This is further complicated by the fact that blockchains are not compatible with each other, i.e., once a company or research group has selected a blockchain for an application, it is locked-in to this technology~\cite{SSFB19}. 

One particular approach to mitigate these issues is to achieve interoperability by allowing blockchains to interact with each other~\cite{underwood16,belchior20}. For instance, atomic swaps~\cite{Herlihy:2018:ACS:3212734.3212736,zakhary2019} can be applied in order to provide interoperability on the level of token exchange without the need of a trusted third party, i.e., a cryptocurrency marketplace. However, even with different proposals for atomic swaps in place today, the basic problem of blockchain interoperability, i.e., that transactions processed in one blockchain never leave the ecosystem of the particular blockchain, is not solved.

This basic characteristic of blockchains also prevents smart contracts from being called in a cross-chain manner, e.g., it is not possible that a smart contract running on a source chain can call a smart contract on a target chain, receive the result from the contract call, and then continue its enactment on the source chain, using this result~\cite{SSFB19}. As a consequence, companies seeking to interact with each other using smart contracts have to agree on a given blockchain in the first place, and changing this afterwards can incur significant costs.

For instance, blockchains and especially smart contracts have been frequently named as facilitators of future supply chain processes, e.g.,~\cite{korpela17,bocek17,petersen18}. In such a scenario, blockchains are used to store data about the progress within a supply chain, and smart contracts can be used to automatically execute payments or to take over the duties of a clearing house. In a fixed setting, the companies involved in a supply chain may agree on a particular blockchain technology and then simply use it without the need for further changes. However, if another company utilizing a different blockchain technology is added to a supply chain, the newcomer has to adapt its existing system and smart contracts in order to participate. It would be preferable if the new company's system and smart contracts were interoperable with the original blockchain technology used by the incumbent supply chain stakeholders.

The missing interoperability between today's blockchain technologies may even decrease technical progress: Choosing novel blockchain technologies enables users to utilize new features and to take advantage of state-of-the-art technology. However, if selecting a new blockchain technology comprises that all necessary data and smart contracts need to be ported from a formerly used blockchain to a new one, users might refrain from testing new technologies, even if this means that novel features remain unavailable~\cite{DBLP:journals/access/BorkowskiSFHS19}.

Therefore, within the work at hand, we propose a framework for invoking smart contracts across blockchains. The presented approach handles the passing of parameters and return values between different blockchains, offers scalability, and allows recursive smart contract calls. We implement our solution for Ethereum-based blockchains and Solidity, but have explicitly foreseen the extension to other blockchain protocols and smart contract languages.

The contributions of this paper are as follows:
\begin{itemize}
	\item We present a framework for cross-chain smart contract invocations. This includes a high-level generic workflow, and the description of low-level choices for implementing this workflow.
	\item We provide an open-source reference implementation in Solidity and evaluate the performance regarding cost and time overheads.
\end{itemize}
The remainder of this paper is structured as follows.
In Section~\ref{relatedwork}, we discuss the state-of-the-art. Afterwards, we present our framework for cross-chain smart contracts in Section~\ref{main_part_1}, followed by discussions in Section~\ref{main_discussions} and an extension of our framework in Section~\ref{sub:extensions}. The evaluation is provided in Section~\ref{evaluation}. Finally, Section~\ref{conclusion} concludes and summarizes this paper.

\section{Related Work}
\label{relatedwork}

\noindent
The number of solutions aiming at cross-blockchain smart contract interactions is quite small. Nevertheless, there are a number of studies~\cite{DBLP:conf/icdcs/0001DX18,ion_stage_blog,qiu19,Herlihy:2018:ACS:3212734.3212736,zakhary2019,DBLP:journals/pvldb/HerlihySL19,fynn20} which are related to the approach presented in the work at hand.

In a seminal discussion, Buterin~\cite{vitalik_crosschain} presents three main approaches for blockchain interoperability, namely \emph{(i)}~notary schemes, where a trusted group claims on chain B that a given event on chain A occurred, \emph{(ii)}~sidechains, where chain B can read chain A via a relay, and \emph{(iii)}~hash-locking, where a secret is hashed by party A and the hash is then published on chain~A by party A and on chain B by party B. Thus, when party A reveals the secret to claim the tokens on chain B, party B learns the secret so that party B can claim the tokens on chain A. The author discusses that hash-locking is not sufficient to achieve portability of any asset such as smart contracts, while the main difference between notary schemes and sidechains is their trust model. For the former, the majority of notaries have to behave honestly, while the latter does not allow sidechains to fail or suffer from ``51\%~attacks''. As we will discuss in Section~\ref{main_part_1}, we also use a notary scheme in our proposed framework.

Jin et al.~\cite{DBLP:conf/icdcs/0001DX18} focus on blockchain interoperability in general by discussing five challenges, namely atomicity, efficiency, security, diversification tolerance and developer friendliness, and propose an event monitor to improve the performance of sidechains.
In addition, the authors state that the biggest problem of cross-chain smart contracts is to connect the runtime environments of two blockchains. Unfortunately, no solution for cross-chain smart contracts is provided.

Ion~\cite{ion_stage_blog} is a work-in-progress implementation of a sidechain to transfer a particular state from one blockchain to another.
The authors focus on validating the state so that it can be used on a different blockchain during the transaction.
However, this concept does not allow to execute a smart contract of a different blockchain. Instead, during the execution of a smart contract, the events issued from a smart contract from a different chain can be accessed.

Liu et al. present HyperService, which facilitates the development of decentralized applications which are executable across the boundaries of different blockchains~\cite{liu19}.
Interaction with other blockchains is done via relays. Notably, the authors provide a blockchain-neutral state model and a high-level programming language which can be used to develop cross-chain decentralized applications. Smart contracts written in different languages can be orchestrated using a program defined in this high-level language. In contrast, we allow the direct invocation of smart contracts without the need to orchestrate them. 
Also, while we apply a validator-based approach to check the correctness of cross-blockchain smart contract invocations (see Section~\ref{main_part_1}), HyperService makes use of an additional ``blockchain of blockchains'' which consolidates the finalized transactions of all participating blockchains and allows to prove actions during cross-chain executions. 

Robinson and Rashed propose a protocol to allow function calls across blockchains which mirrors the goals of our work~\cite{robinson20}. For this, the authors apply Ethereum events, which are emitted by trusted smart contracts on each participating blockchain. Notably, the approach requires the predefinition of signers and administrators, with the administrators being responsible for adding further blockchains (via trusted contracts). In contrast, the intermediaries and validators used in our work do not have to be predefined. Nevertheless, the work by Robinson and Rashed comes closest to the work at hand.

Fynn et al.~\cite{fynn20} follow a distinct approach to blockchain interoperability by providing the means to migrate a smart contract from one blockchain to another. While this is an interesting approach, the authors do not regard that smart contracts on different blockchains may interact.

Qiu et al.\cite{qiu19} present ChainIDE, which is an IDE specifically aiming at smart contract developer support. Among other aspects, the authors also discuss the need to support developers when implementing smart contracts for different blockchain protocols. However, interactions between smart contracts running on different blockchains are not regarded.

In general, the related work primarily focuses on transferring tokens from one blockchain to another. For instance, Metronome\cite{metronome} proposes an import/export system for transferring tokens between two blockchains. 
For this, a user has to ``destroy'' the tokens on the source chain to gain a proof-of-exit receipt, which can be used on the target chain to reclaim the tokens. 
Metronome utilizes validators, which verify whether the receipt is correct or not.

Borkowski et al. discuss the Deterministic Cross-Blockchain Token Transfers (DeXTT) protocol~\cite{DBLP:journals/access/BorkowskiSFHS19}. DeXTT allows to synchronize token balances across an arbitrary number of chains. Thus, DeXTT ensures that tokens remain available even if one of the chains is not operating any longer. For this, the DeXTT protocol utilizes intermediaries, called witnesses, which verify and broadcast transactions to all participating blockchains.

Cosmos~\cite{cosmos_whitepaper} uses a blockchain, called hub, which is connected to independent blockchains, called zones, to transfer tokens as packets between zones through an inter-blockchain communication protocol. 
In this approach, the receiving chain has to prove a packet by keeping track of the block headers of the sender chain.

Polkadot~\cite{polkadot_whitepaper} uses a relay chain to distribute transactions to multiple ``worker chains'' called parachains, which process the transactions. 
The solution allows to include external blockchains like Ethereum via bridges. Validators forward transactions to the external blockchains, and blockchain-specific properties such as events and Merkle proofs back to Polkadot.
The validity of a block is proven by two-thirds majority and by challenges that can be submitted after the vote in a certain time period to prove the invalidity of the header data of a block. 
Notably, the relay chain does itself not support smart contracts, but this functionality might be added by parachains in the future.

Interledger~\cite{interledger} provides a protocol for routing payments, e.g.,~tokens or fiat currency, between different payment networks, e.g., blockchains. 
They establish a payment channel between two parties off-chain and exchange token claims, which are published after a predetermined time on an on-chain network for distributing the funds appropriately between the involved parties.

Recently, some further approaches to cross-blockchain transactions have been proposed, e.g.,~\cite{zhao20,wang20}. While such approaches provide important insights on how to realize
transactions across two or more blockchains, they do not implicitly aim at smart contract interoperability across different blockchains.

As it can be seen from the discussion, some approaches aiming at blockchain interoperability have been presented so far. However, smart contract interoperability has~--~to the best of our knowledge~--~ not been extensively discussed so far.

\section{Cross-Chain Smart Contract Invocations}
\label{main_part_1}

\noindent
In this section, we introduce our framework for cross-chain smart contract interactions. We first define the basic terminology used within this paper (Section~\ref{sub:terminology}). In Section~\ref{sub:sequence}, we  discuss the different phases which occur in cross-chain smart contract interactions. In Sections~\ref{register_phase} through \ref{sub:finalization_phase}, we then present these phases in detail. 

\subsection{Terminology}
\label{sub:terminology}

\noindent
Cross-chain smart contract interactions are based on the assumption that a user or smart contract calls a function of an account on a different blockchain and potentially receives a response (e.g., some return values) from the called account. We denominate the calling entity as the \emph{caller}, while the account which is called is the \emph{callee}. The \emph{source chain} contains the account of the caller, while the \emph{target chain} contains the account of the callee. 

As we will discuss in the following sections, in our approach, two smart contracts are needed to enable smart contract interactions, i.e., to call a method on a target chain, and to return the results to the caller: The \emph{distribution contract} on the source chain and the \emph{invocation contract} on the target chain. Apart from the caller and the callee, further users are involved in our scenario, namely the \emph{intermediaries}, who play the role of brokers and transfer information between the source and the target chain, and the \emph{validators}, who validate the information passed by the intermediaries.

\subsection{Overview}
\label{sub:sequence}

\begin{figure}[t]
	\centering
	\includegraphics[width=\columnwidth]{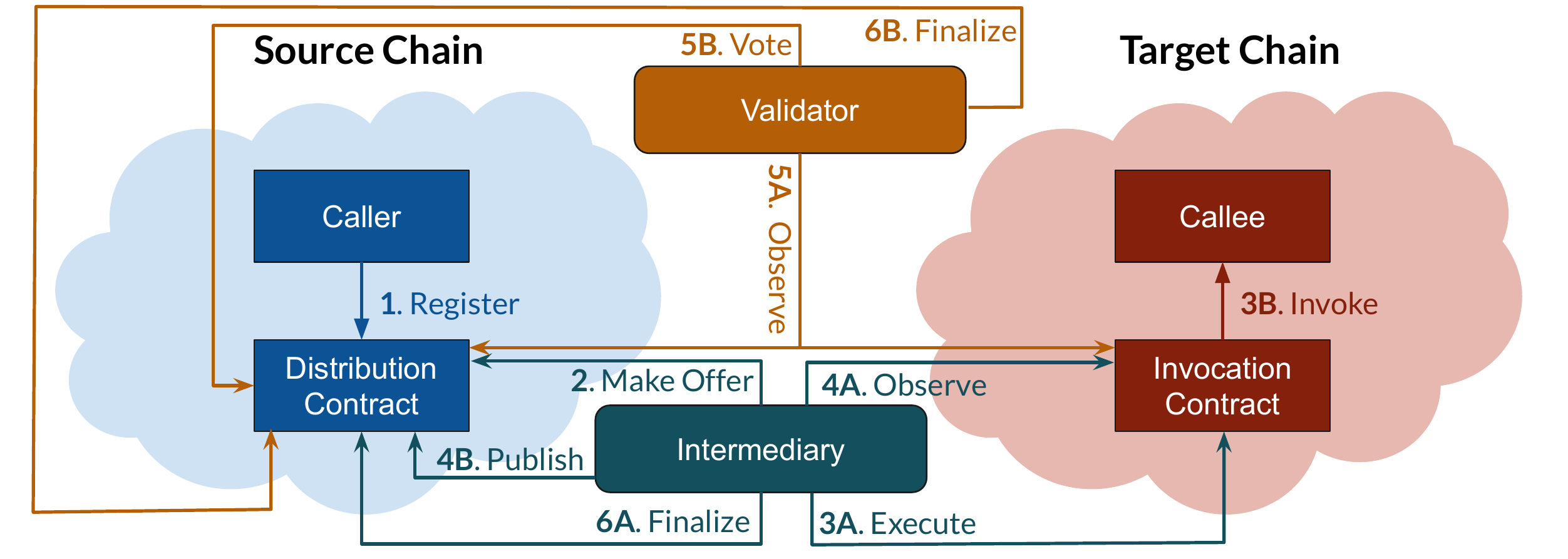}
	\caption{Interactions between caller, distribution contract, intermediary, invocation contract, callee and validator}
	\label{fig:design_architecture}
\end{figure}

\noindent
Figure~\ref{fig:design_architecture} summarizes the workflow of a cross-chain smart contract interaction following our approach. Similar to recent literature on cross-token projects, e.g., \cite{DBLP:journals/access/BorkowskiSFHS19,metronome,polkadot_whitepaper}, our work follows a broker-based approach, i.e., third parties are used to interconnect the source and the target chains. 
We have chosen this approach in order to provide a simple way to call a contract account on the target chain. This would not be possible by using a relay chain, since each transaction then would have to be started by an external account.

\smallskip
\noindent
The overall workflow consists of the following phases:
\begin{enumerate}
	\item \textbf{Register phase}. In order to invoke a contract on the target chain, the caller \emph{registers} the invocation at the distribution contract on the source chain. 
	The distribution contract saves the metadata of the call, e.g.,~which blockchain is the target chain, and which parameters are provided to the callee, and announces the metadata via an event to the intermediaries.
	An event informs listening applications about the current state of the contract. Such events are stored with their arguments, for example, by using the logging functionality of the blockchain and can therefore be inaccessible within contracts. 
	The distribution contract manages the complete process of the cross-chain interaction. In contrast, the invocation contract on the target chain is required to save the return values, since these values are not observable by an external account, i.e., an account hosted on the source chain (see Section~\ref{register_phase}).
	
	\item \textbf{Offer phase}. As pointed out above, intermediaries are used to broker information between the source and the target chain, i.e., between the distribution contract and the invocation contract. 
	An intermediary is a third party which has accounts on both blockchains and can interact with both technologies by listening for state changes, e.g.,~events, and publishing transactions.
	Once a register event has been published, parties can \emph{make an offer} to become the intermediary for a particular cross-chain smart contract interaction. Parties are monetarily incentivized to become an intermediary (see Section~\ref{offer_phase}).
	
	\item \textbf{Execution phase}. Once an intermediary has been selected, it can \emph{execute} the transaction on the target chain. This is done through the invocation contract, which in turn \emph{invokes} the callee (see Section~\ref{sub:execution}).
	
	\item \textbf{Forwarding phase}. After the smart contract on the target chain has been invoked, the intermediary \emph{observes} the blocks on the target chain in order to extract the results of the smart contract invocation. 
	The intermediary takes the results from the target chain and \emph{publishes} them by invoking the distribution contract on the source chain. Thus, the data is now available on the source chain and therefore to the caller (see Section~\ref{sub:forwarding}).
	
	\item \textbf{Verification phase}. In order to protect callers from using incorrectly published data, the return values have to be approved. 
	Therefore, validators \emph{observe} both the results of the invocation on the target chain and the published return values on the source chain. To conclude whether the published values are correct or not, the validators \emph{vote} on their correctness (see Section~\ref{sub:verification}).
	
	\item \textbf{Finalization phase}. In case of a positive vote, the transaction is \emph{finalized} by the intermediary, otherwise by the validator. In this step, an honest intermediary and the validators get a reward for their participation (see Section~\ref{sub:finalization_phase}).
\end{enumerate}

\noindent
The discussion of the following phases focuses on Ethereum-based smart contracts. This is done to keep the discussion based on one concrete case, and since different concepts in the handling of transaction costs and interactions with smart contracts have to be discussed individually per blockchain.
However, these individual concepts do not have any direct consequences on the basic workflow of our approach. We discuss in Section~\ref{sub:support_other_blockchains} how our work could be ported to other blockchains protocols.

\subsection{Register Phase}
\label{register_phase}

\noindent
On a local chain, a smart contract invocation requires the definition of \emph{(i)}~the address of the contract, \emph{(ii)}~the method to be called,  \emph{(iii)}~the parameter values of the method, \emph{(iv)}~a maximum number of computational steps (startgas), \emph{(v)}~the gas price per computational step (gasprice), and \emph{(vi)}~the number of tokens to be transferred to the callee~\cite{ethereum_whitepaper}.

Traditionally, a smart contract can be called only within its own chain. For cross-chain smart contract interactions, this means that a caller on the source chain cannot call a smart contract on a target chain directly. Hence, we have to transfer and adapt the discussed requirements for cross-chain smart contract invocations. 

As mentioned in Section~\ref{sub:sequence}, we use intermediaries to invoke a smart contract on the target chain.
Therefore, we have to announce a cross-chain smart contract call with the required data, i.e.,~\emph{(i)-(vi)}, to potential intermediaries.
While the data could be transferred off-chain or on-chain to the potential intermediaries, we use the already provided infrastructure by the blockchain and publish the data directly on the source chain.
For this, we implement the distribution contract, which assigns each registration a unique invocation id, accepts all relevant parameters (described in the next paragraphs), stores them and announces an event for the potential intermediaries.

For the cross-chain invocation of a smart contract, the above mentioned definitions \emph{(i)-(vi)} need to be extended. Most importantly, in a cross-chain setting, information is necessary on which chain the invocation should take place. For this, we add an additional blockchain identifier to the invocation information. As is the case in intra-chain settings, the identification data for a smart contract, i.e., \emph{(i)-(iii)} \emph{plus} the additional identifier needs to be provided by the caller.

Function-specific data, i.e., the startgas (\emph{iv}) also needs to be provided by the caller. However, it should be noted that function-specific data depends on the behavior of the called smart contract and the used parameters. This data cannot be estimated without knowing the logic of the smart contract. As in intra-chain settings, this data needs to be  defined by the caller as well.

Finally, target chain-specific data \emph{(v)-(vi)} needs to be defined. In our current approach, we focus on smart contract interactions and therefore omit token transfers. 

For the definition of the gas price, various possibilities exist, which we discuss in the following paragraphs:

\smallskip\noindent
\textit{Caller sets Gas Price.}
The caller determines the gas price of the target chain when registering the transaction.
Usually, the gas price of the target chain depends on the workload of the blockchain and is therefore a dynamic value which has to be determined per registration.
Since the caller may not have access to the dynamic information about the target chain, e.g., contract-based accounts, this approach should not be used in cross-chain invocations.

\smallskip\noindent
\textit{Contract sets Gas Price.}
As a second possibility, the distribution \emph{or} the invocation contract can be used to store and manage the gas price, which is then used by the intermediary for the transaction.
In order to adjust the price to the current workload of the blockchain, participants (e.g., caller, validator, intermediary) can vote on a change in the gas price.
However, this requires a large number of additional transactions to continuously adjust the gas price in the contracts and is therefore not suitable.

\smallskip\noindent
\textit{Intermediary sets Gas.}
Per se, the intermediary has access to both blockchains. Therefore, it is aware of the current gas price of the target chain.
In order for the caller to pay the transaction costs on the target chain, the gas price must be converted by the potential intermediary from the target chain to the source chain using an exchange rate.
However, the intermediary aims for profit and since there would be no control institution that can verify, without consensus finding, whether the chosen exchange rate is acceptable, the intermediary would set the gas price to a maximum value, so that a large amount of gas of the caller is used.

\smallskip\noindent
\textit{Gas Price Competition.}
As it can be seen, the possibilities to choose the gas price discussed so-far have individual weaknesses. 
Therefore, we implement a different solution, which we call the \emph{Gas Price Competition}. 
Following this approach, instead of letting an intermediary set the gas price directly, the potential intermediaries propose a gas price in the currency of the source chain during the offer phase (see Section~\ref{offer_phase}). 
This leads to a competition for the lowest offer, as the candidate with the lowest gas price wins the competition and therefore becomes the current intermediary.
On the one hand, the respective intermediary itself has to ensure that the ratio between the gas price submitted in the offer on the source chain and the actual gas price on the target chain is not below the current exchange rate, since otherwise the offer is not lucrative for the intermediary, but on the other hand if the ratio is higher, more profit can be made.
Due to the competition between the potential intermediaries, it can be expected that the value settles to the exchange value.
However, if there is no competition between the potential intermediaries, or they find a way to fix a gas price among themselves, the caller needs to be protected against excessive gas prices. For this, we allow the definition of a maximum gas price.

As is the case in intra-chain settings, the caller also has to pay a transaction fee for the invocation of smart contracts on a target chain. In addition, the intermediary and the validators have to be monetarily rewarded in order to incentivize them to participate. For paying these fees, the caller has to deposit tokens at the distribution contract. This deposition needs to take place during the register phase. Else, there is a risk that the caller will not pay the fees to the intermediary and the validators after the invocation has taken place.

The minimum amount of tokens to be deposited can be calculated by
\begin{equation}
	\begin{aligned}
		\textrm{max~gas price} \cdot \textrm{startgas} + \textrm{fees} \leq \textrm{deposited tokens}
	\end{aligned}
\end{equation}

\noindent
To summarize the register phase, the caller registers the cross-chain smart contract invocation with the identification data \emph{(i-iii)} \emph{plus} the \emph{identifier} of the target chain, a startgas \emph{(iv)}, and a maximum gas price, and deposits tokens at the distribution contract for paying transaction data and rewards. The actual gas price \emph{(v)} is proposed by the potential intermediaries during the offer phase (see below).

\subsection{Offer Phase}
\label{offer_phase}

\noindent
After the caller has registered its cross-chain smart contract call at the distribution contract, the latter publishes an event on the source chain, announcing the begin of the offer phase.
Potential intermediaries listen for each retrieved block of the source chain for such events and will submit zero or more offers regarding an event to the distribution contract. As pointed out in the previous section, the submitted offers have to include a gas price, which is used at the end of the offer phase to determine the winner of the competition.

There exist various possibilities to end the offer phase, which we discuss in the following paragraphs.

\smallskip\noindent
\textit{Caller-based.}
The caller sends a transaction to end the offer phase.
This has the advantage that the caller decides whether the proposed offer meets its wishes.
However, if the caller itself is a smart contract, it cannot send the transaction without being invoked by an external account.
Therefore, this option is not investigated further.

\smallskip\noindent
\textit{Random-based.}
The phase ends pseudo-randomly after an offer has been submitted, either by using insecure blockchain parameters or external services~\cite{randomnumbereth}.
In the worst case, either the offer phase never ends because the condition is not fulfilled or the phase can end after the first offer, which prevents an offer competition.
This approach is not suitable for time-critical calls where a maximum duration needs to be guaranteed.

\smallskip\noindent
\textit{Time-based.}
In order to guarantee a maximum execution time, we suggest to use a time-based approach, where the offer phase ends after a predefined time, e.g., defined by the caller or chosen by the distribution contract, and the winner is selected implicitly without any additional transaction.
In case that no offer has been submitted in time, an additional handling is required to return the deposited tokens to the caller.
For this, we suggest to abort the cross-chain call and directly jump to the finalization phase, where the call has to be finalized by the caller (see Section~\ref{sub:finalization_phase}).
Note that the time-based approach can be combined with the random-based one by using \textit{soft limits} or \textit{hard limits}, i.e.,~the offer phase remains active at least until the soft limit, it can randomly end between the soft and the hard limit and will definitely end at the hard limit.
While this may increase the incentives for submitting better offers, there is no technical advantage. It increases the complexity and loses the deterministic properties.

As mentioned, the winner of the offer phase is determined based on the lowest submitted gas price.
In the case of a tie, we make use of an algorithm introduced in DeXTT~\cite{DBLP:journals/access/BorkowskiSFHS19}, where the winner is determined on the basis of the lowest hash value, consisting of the invocation id, the address of the intermediary, and the gas price.
This guarantees that the winner is selected in a deterministic way without any additional transaction and does not favor intermediaries with a faster network connection.

\subsection{Execution Phase}
\label{sub:execution}

\noindent
Once an intermediary has won the offer, it has to process the smart contract call on the target chain.
For this, the intermediary submits the data received from the distribution contract on the source chain to the invocation contract on the target chain, which forwards the transaction to the callee. 
The invocation contract is required for two reasons:
First, an external account cannot receive the return values of a function call, hence, a contract has to store it.
Second, the invocation contract simplifies the access of the parameters during the verification process of the validators, which is discussed in detail in Section~\ref{sub:verification}.

Apart from storing the input and output data, the invocation contract assigns each execution a unique execution id for tracking purposes and manages the execution status.
This is important for extended functionality, such as recursive cross-chain invocations, which are discussed in Section~\ref{sub:extensions}.

Note that in this step the intermediary has to pay the transaction costs in advance and will get them reimbursed in the finalization phase (see Section~\ref{sub:finalization_phase}). 
Otherwise, transaction costs could be collected without following the next phases to steal the tokens from the caller.

\subsection{Forwarding Phase}
\label{sub:forwarding}

\noindent
After the smart contract invocation on the target chain has been successfully processed, the intermediary has to forward the return values to the source chain.
For this purpose, the intermediary observes the blockchain and waits for an event announcing the completion of the call.
Once such an event is observed, the intermediary extracts the return values from the invocation contract and publishes it on the distribution contract of the source chain by passing the value as argument of a function. 

Although the value is now available on the source chain and can be used by the caller, this does not guarantee that the value is correct.
For example, the intermediary could publish random data without interacting with the invocation contract at all. 

\subsection{Verification Phase}
\label{sub:verification}

\noindent
To avoid incorrect data publication, the result published by the intermediary is verified for correctness.
For this purpose, validators are used which compare the data stored in the distribution contract of the source chain with the data stored in the invocation contract of the target chain. 
This data includes:
\begin{itemize}
	\item Execution of the correct callee (blockchain identifier, address, method, parameters)
	\item Publication of the correct result (status, value) from the correct transaction (invocation id, execution id)
	\item Correct execution boundaries (startgas, actual ``steps'')
\end{itemize}

\noindent
In order to determine whether the cross-chain call was correct, the validators vote among three different outcomes: \emph{(i)} the call was invalid, \emph{(ii)} the call was valid and \emph{(iii)} the call was valid, but the actual number of ``steps'' does not match, i.e.,~the intermediary submitted more steps than were actually needed.
The third option is introduced to avoid resubmitting the complete transaction if the transaction was executed correctly but the intermediary cheats on the number of computational steps in order to receive more tokens. These ``step votes'' are not done separately, but are part of the normal voting procedure.

As in the offer phase, the voting ends after a predefined time.
Then, the positive (call valid) and negative (call invalid) votes are compared.
If the vote is positive, the ``step votes'' are compared.

The reward is distributed to only one validator among the participants in the winning group.
Similar to the offer phase, the winner is chosen by selecting the validator with the lowest hash, which is for this purpose defined as the hash of the unique voting id and the validator address. For a complete discussion on the deterministic selection of the reward recipient, we refer to \cite{DBLP:journals/access/BorkowskiSFHS19}.

\subsection{Finalization Phase}
\label{sub:finalization_phase}

\noindent
After result verification is completed, the cross-chain smart contract call waits for finalization on the distribution contract. 
Finalization marks the call as completed and distributes the rewards. 

Depending on the voting result, the tokens are distributed differently: If the voting result is negative or the comparison of the computational steps failed, the entire transaction value with the exception of the result verification reward is reimbursed to the caller. This means that the intermediary only receives the reward and gets reimbursed the transaction costs if its behavior was honest over the whole execution and it has not tried to cheat the system.
Therefore, it is likely that the caller or the validators will finalize the invocation due to their respective incentives.

If the voting result is positive, the intermediary gets reimbursed the transaction costs plus its reward, the winning validator of the voting will receive its reward and the remaining tokens will be refunded to the caller.
Since the intermediary receives most of the tokens, it is assumed that the transaction will be executed by the intermediary, although it can be called by anyone else, such as the caller or the validator.

As mentioned in Section~\ref{offer_phase}, if no offer has been submitted, the call has to be finalized as well so that the deposited tokens are released to the caller. 
Since there is no voting and no cross-chain execution, neither the intermediary nor the validator will finalize the cross-chain call because they receive no reward.
Therefore, only the caller will most likely execute this function in order to get its deposit back.

Upon completion of this phase, the caller can access the status and the returned value by providing the invocation id to the distribution contract.

\section{Discussion}
\label{main_discussions}
\noindent
In this section, we discuss misbehavior, i.e., what is done in case an intermediary is behaving in a malicious way (see Section~\ref{misbehavior}), the voting mechanism necessary for result verification and misbehavior detection (see Section~\ref{voting}), how the intermediaries and validators can be selected (see Section~\ref{sub:selection_intermediary_validator}), why we only allow the execution from one intermediary (see Section~\ref{sub:selection_number_of_intermediaries}),  and how this approach can be applied to further blockchain protocols (see Section~\ref{sub:support_other_blockchains}).

\subsection{Penalizing Misbehavior }
\label{misbehavior}

\noindent
One particular case of misbehavior occurs when the intermediary gives up its reward and does not execute the transaction in its entirety.
Then, the deposit of the caller would be locked forever.
To solve this problem, two aspects have to be considered:
First, the intermediary should be penalized and second, the tokens have to be released.

\smallskip\noindent
\textit{Penalty fees for intermediary.}
To penalize the non-execution of a cross-chain smart contract call, we suggest that intermediaries have to deposit tokens in a depot with which the misbehavior can be compensated.
A pre-defined amount of tokens are locked from the deposit on submission of an offer and become available for further use once the cross-chain call has reached the final state or a better offer has been made.

Each time the intermediary does not execute the transaction, the locked tokens are used to pay a penalty for violating the ``execution contract''.
Therefore, we suggest as a minimum the same amount of tokens as fees for the intermediary as the deposited fees by the caller. Thus, the intermediary could pay as a penalty the same amount of tokens for the validation process (see next paragraph) and to the caller for its misbehavior (reversal of reward).

\smallskip\noindent
\textit{Release locked tokens.}
In order to release the locked tokens to the caller, a method for detecting misbehavior has to be defined.
For this, we propose a combination of a time-based and a voting-based detection method to protect against misuse of voting attempts. Thus, voting can only be started after a certain period of time. For example, this could be a fixed time period (e.g., 4 hours), a fixed amount of blocks (e.g., 1,000), or be explicitly defined by the caller.

In this period of time, the intermediary should have enough time to publish the result, or in case of a recursive cross-chain call (see Section~\ref{sub:extensions}), to register the status of the transaction on the source chain so that validators can observe if the cross-chain call is still in process.
After this time span, each validator can start to vote for a fraudulent cross-chain call, i.e.,~a call that has not been (completely) processed by the intermediary. 

To encourage validators to participate in the voting, a voting reward is paid with the same distribution criteria as in Section~\ref{sub:verification}.
However, this time the fee is either paid by the intermediary (if it is dishonest) or by the voting initiator. 
The validator has to deposit tokens similar to the intermediary before starting this voting.
This has the positive side effect of a ``protective charge'', i.e., that validators will usually avoid to start invalid votings.
\subsection{Voting Mechanism}
\label{voting}

\noindent
As mentioned in Section~\ref{sub:verification}, we discuss the voting mechanism of our framework in this section.
For both votings (result verification and misbehavior detection), we use a majority voting system, i.e.,~the result with more than 50\% of votes is assumed to be correct.
This voting procedure should ensure that an honest party does not lose any tokens under the assumption that at least 51\% of the validators behave honestly.
However, the probability of voting decreases exponentially per voting option because a validator will only submit its voting when it can win. Since we make use of a hash-based winner selection, the hash of a validator has to be lower than the current winner. 
Therefore, in practice, the worst case is a low hash, e.g., zero, for the first vote for a voting option, resulting in no further votes for this option.

Since the source chain does not know the truth value of the target chain and can only judge by the majority of votes regardless of the correct result, validators have no incentive to behave honestly. Naturally, validators will aim for profit and always vote for the current majority if their own signature is lower.
Therefore, it is important to monitor the validators.
This can be achieved, for example, with a delegated Proof of Stake protocol~\cite{dpos}, where participants will only nominate honest validators.
We want to point out that the incentives to follow the majority-voting are not specific to our solution, but can also be found in a similar way in other validator-based approaches using majority voting, e.g.,~\cite{metronome, polkadot_whitepaper}.

One possibility to increase the incentive of voting correctly is the usage of zero-knowledge proofs~\cite{DBLP:journals/iacr/PanjaR18} to hide the votes until the end of the voting phase.
Since this paper focuses on a generic framework to cross-chain smart contract interaction, we leave such further improvements to our future work.

We further discuss the security implications of relying on (dishonest) third parties. 
In case that only the intermediary is dishonest, the introduced countermeasures are executed, which leads to the loss of the rewards plus an additional penalty fee for the intermediary. In addition, the caller has to resubmit the transaction leading to an additional time-delay. 
In case the validators are dishonest, they either act in favor of the caller or in favor of the intermediary. In the former case, a valid transaction is marked as invalid, implying the same consequences as above, but this time the intermediary illegally looses the reward. In the latter case, they mark an invalid transaction as valid. This results in the caller being not aware of an invalid result and in the worst case carrying over an invalid result to further transactions. In addition, the caller has to pay for the transaction.

\subsection{Selection of Intermediaries and Validators}
\label{sub:selection_intermediary_validator}
\noindent
As discussed in Section~\ref{main_part_1}, we use intermediaries and validators to transfer and verify the information between the source and the target chain. 
To ensure that intermediaries and validators cannot launch any Sybil attacks against the framework, they have to be selected by the participants or by authorization. While we established this via authorization in our proof of concept, we refer to use for the selection process in a decentralized setting an established mechanism such as a Sybil attack-resistant delegated Proof of Reputation protocol~\cite{10.1145/3343147.3343160} where participants can vote for intermediaries and validators.
Such voting protocol has to ensure that intermediaries and validators are not the same nodes, otherwise the intermediaries can verify their own transactions and defraud the system.
In addition, voting-based protocols allow participants to stop voting for validators and especially intermediaries which do not submit, or publish wrong information.

\subsection{Number of Intermediaries}
\label{sub:selection_number_of_intermediaries}
\smallskip\noindent
While we support multiple validators to be part of one cross-blockchain call, we have limited the transaction execution to only one intermediary. Initially, this could be considered to be a disadvantage due to decentralization aspects, but indeed, this approach features several advantages:
First, transaction execution by multiple intermediaries would result in a linear increase in gas costs leading to a cost exposure. Since transactions can become very expensive, the caller must either reimburse the transaction costs to all intermediaries, which is a cost exposure for the caller, or the intermediaries do not get the transaction refunded which will lower the number of participants in the process/offering phase. 
Second, some transactions return different values per execution, for example, a UUID generator. 
For this reason, we limit the number of intermediaries to exactly one and impose penalties in case of misconduct to reduce the number of invalid cross-chain calls.

\subsection{Support of further Blockchain Protocols}
\label{sub:support_other_blockchains}
\noindent
As mentioned in Section~\ref{sub:sequence}, the presented approach has been implemented for Ethereum-based blockchains, and the actual adaptation of the workflow depends on the semantics imposed by the blockchain technologies and their smart contract languages.
We require as a minimum smart contract functionality the ability to store data, perform arithmetic operations, and enforce conditions. 
Notably, our implementation for the intermediary and validator requires knowledge of the functionality of the blockchains and has to be extended for further blockchains, for example adapted to blockchain-specific interfaces and behavior in order to process and validate transactions correctly (read gas price, event and storage access, etc.). In the following, we show how the support for a different technology can be added, using EOS as an example. 

EOS is a blockchain technology that creates an operating system-like construct upon which applications can be built, which fundamentally differs from the concept of Ethereum. 
In detail, EOS smart contracts deviate from Ethereum smart contracts in two relevant points. 
First, they do not support events and second, they have no transaction costs. 
For the former, we change the strategy to a polling-based approach, where the distribution and invocation contracts are checked for changes after each block, while for transactions costs the gas parameters are set to zero when an EOS contract is to be invoked. 
If an Ethereum contract is called from EOS, nothing has to be changed, i.e., transaction costs remain as defined, but are paid with EOS tokens, which is the source chain currency.

It should be noted that the chosen implementation of Ethereum-based blockchains sufficiently allows us to evaluate the basic functionality of our approach to smart contract interoperability. Implementing the proposed approach first for this technology has also the benefit that Ethereum-based blockchains (e.g., Ethereum, Ethereum Classic, Ubiq, WanChain) today provide the most popular solutions for decentralized applications and digital assets~\cite{cai2018decentralized, diangelo2019survey}. Cross-chain smart contract invocations for Ethereum-based blockchains can thus enhance the utility of a majority of available assets. 
Notably, our implementation can also be applied to sidechains for Ethereum. Sidechains are a means to increase the transaction throughput in a blockchain network. For this, certain transactions are offloaded to the sidechain~\cite{back2014enabling}. 

Last but not least, we note again that the proposed approach does not pose any restrictions on the concrete means of cross-blockchain interaction. As long as a blockchain provides sufficient scripting capabilities to implement the presented concepts, the solution can be adopted beyond Ethereum-based blockchains, as described above for EOS.

\section{Extensions}
\label{sub:extensions}

\noindent
In this section, we provide two important extensions to the basic workflow between source and target chains presented in Section~\ref{main_part_1}.

\subsection{Continuation on Source Chain}
Initially, the caller does not know when a cross-chain smart contract is finished.
Therefore, the caller has to observe the blockchain to see if the cross-chain call is finalized to use the value in subsequent transactions.
In order to reduce the caller's work, it is advisable to automatically execute the subsequent transaction once the cross-chain call is completed, especially if the caller is a contract-based account that cannot watch the blockchain for results without being invoked.

Therefore, we extend the registration process presented in Section~\ref{register_phase} with additional parameters (startgas, gas price and method name for the callback function) to let the caller define a callback function, which is executed during the finalization phase.
As a consequence, the formula for depositing tokens introduced in Section~\ref{register_phase} has to be adjusted to include the deposition of the callback function:
\begin{equation}
	\begin{aligned}
		\textrm{max~gas price} \cdot & \textrm{startgas} + \textrm{fees}~+ \\ \textbf{cb. startgas} & \cdot \textbf{cb. gas price} \leq \textrm{deposited tokens}
	\end{aligned}
\end{equation}

\noindent where \textit{cb. startgas} is the maximum number of computational steps and \textit{cb. gas price} is the price paid per computational step of the callback function.

\subsection{Recursive Smart Contract Calls}
\noindent
In some cases, the return values of the callee depend on an additional cross-chain call, which creates a chain of cross-chain calls.
For this, we extend the execution phase defined in Section~\ref{sub:execution} by adding an additional status flag (pending) to the invocation contract, which informs the intermediary that the result is not ready for the source chain due to a recursive cross-chain call. 
Once this call is finished, the flag is updated either to \textit{pending} or \textit{completed} depending on whether an additional call is required.
The process continues as defined in Section~\ref{sub:forwarding} by transferring the return values to the source chain when the flag is set to completed.

\begin{figure}[t]
	\centering
	\includegraphics[width=.85\columnwidth]{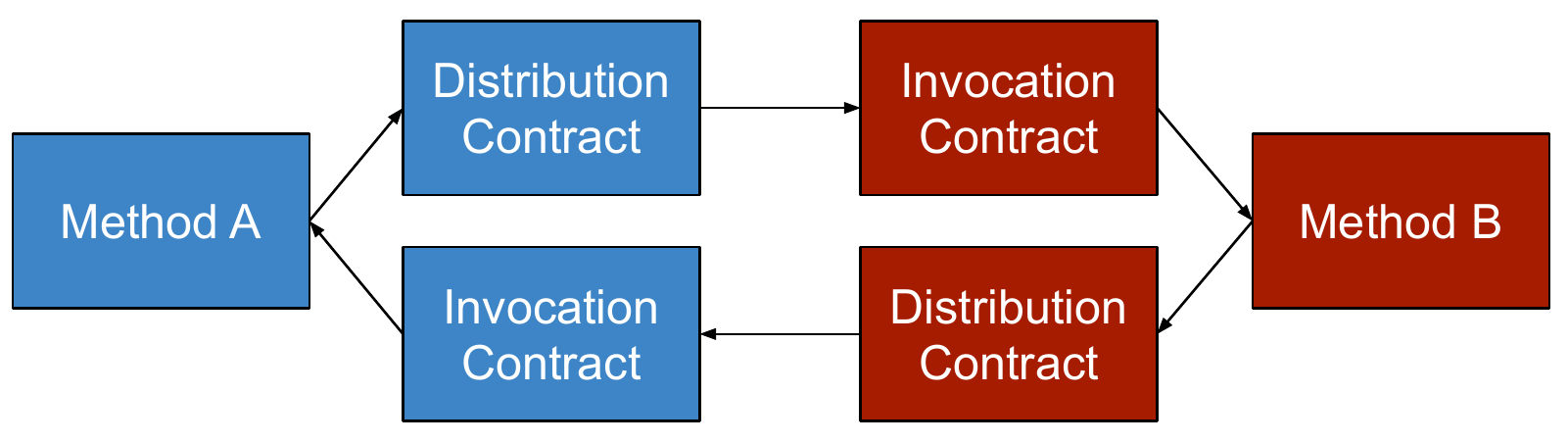}
	\caption{Recursive Execution Attack}
	\label{fig:RecExecAttack}
\end{figure}

This procedure raises the risk of a possible recursive execution attack, where at least two chains call each other unboundedly (visualized in Figure~\ref{fig:RecExecAttack}). 
An attacker may create such a scenario to prevent an intermediary from ever finishing the given transaction and getting its deposit tokens for the transaction back.

One possible solution against this attack is to limit the maximum number of recursive chain calls.
For this, the intermediary transfers the current number of recursive chain calls to the other blockchain, which is verified by validators to prevent attacker-controlled intermediaries from changing the number.
In our current solution, we have not implemented any check against this form of attack since in the basic workflow it only harms the intermediary but does not have any real advantage for the attacker.

\section{Evaluation}
\label{evaluation}
\noindent
In this section, we evaluate the implemented approach regarding time and costs.
We implemented the contracts in Solidity, the primary language for Ethereum, and the logic for the intermediaries and the validators in node.js using web3.js. The source code is publicly available\footnote{
	\url{https://github.com/markusnissl/cross-chain-smartcontracts}
}.
For the setup, we use the Rinkeby test network, where we can observe whether the transaction pool or the network latency has an impact on the duration of the cross-chain call.
The blockchain is initialized with the developed distribution and invocation contract with enough funds for all parties and with a test contract, which provides getter and setter methods for different variable types, e.g.,~int, string, byte.

\subsection{Cost Evaluation}

\noindent
For evaluating the costs of various functions, we compare the implemented cross-chain smart contract call with executing the call manually on the target chain.
For manual invocation, we assume that getter functions are free of charge since no transaction is executed when viewing the data and only setters have to be paid. 
Thus, for both scenarios (setting or getting data), only one setter call is required (either on the target chain for setting or on the source chain for returning the data).
In the proposed solution, we make use of the callback function for returning the data in the finalization phase.

Since the required steps depend on the behavior of the function, the costs of each function are summarized in Table~\ref{tbl:evaluation_gas_framework}. 
The result shows significant costs in all steps, especially when dealing with byte data types, i.e.,~parameters or return values. 
The execution part, which is equivalent to the setter function on the target chain, shows an overhead by a factor of 9.
This can be explained by high costs for creating or modifying storage data.
Overall, the gas costs are about 40 times higher, when taking only one intermediary, one validator and no rewards into account.

Although the values are high, the proposed implementation offers a novel framework for cross-chain smart contract invocations. Thanks to intermediaries and validators, we do not leave the blockchain world and the execution is verified, which is not possible by manually importing the values or if using a non-blockchain third party for the interactions.

\begin{table}
	\centering
	\caption{Required Amount of Gas per Transaction}
	\label{tbl:evaluation_gas_framework}
	\begin{tabular}{l|l}
		Operation & Required Steps\\
		\hline
		Registration w/o callback (\textbf{S}etter) & 362,967-388,568 \\
		Registration with callback (\textbf{G}etter) & 354,325-354,389 \\
		Offer Request (first/better/worse) & 72,448 / 49,011  / 26,131\\
		Target Chain Execution (\textbf{S}/\textbf{G}) & 290,667-331,818 / 265,403-306,348 \\
		Forwarding (\textbf{S}/\textbf{G}) & 309,081 / 344,445-384,982 \\
		Voting (first/other vote) & 110,267 / 60,678 \\
		Finalization without callback & 93,733 \\
		Finalization with callback & 111,114-131,353 \\
		\hline
		Manual calls (\textbf{S}/\textbf{G}) & 26,983-33,022 / 0 \\
	\end{tabular}
\end{table}

\subsection{Time Evaluation}

\noindent
In this section, we compare the time spent in the proposed implementation with the manual execution for a simple cross-chain call.
For the proposed implementation, we execute the transaction three times for various configurations of $\mathit{waitingBlocks}$  and $\mathit{blocksPerPhase}$ and calculate the average block difference between the register and the finalization phase.
The parameter $\mathit{waitingBlocks}$ is used after the registration, the offer, the execution, the result forwarding and the validation phase to wait for consistency in the blockchain, i.e., to avoid small forks, while the $\mathit{blocksPerPhase}$ parameter defines the time for the offer and voting phase.

To calculate the time for a manual execution, we use the $\mathit{waitingBlocks}$ parameter, since it represents the time we have to wait until the value is considered valid in the target chain (independently of reading or writing).
For the source chain, we do not have to add an additional $\mathit{waitingBlocks}$ period for importing the value, since this period is outside the scope of the framework, i.e.,~after the finalization phase.

The results are shown in Table~\ref{tbl:evaluation_time_optimal_rinkaby}, where \emph{m} denotes the manual execution time (in blocks and minutes) and the other values list first the optimal time for the parameters and then the average time for the Rinkeby test network in blocks and rounded to minutes assuming that 4 blocks are mined on average per minute.
The optimal number is calculated by $5 \cdot \textit{waitingBlocks} + 2 \cdot \textit{blocksPerPhase} + 4$, where the factor 5 and 2 describe the number of uses of the parameters in the framework and the summand $4$ is derived from the one-block waiting times until the framework receives the next transaction to continue, i.e.,~before execution, result forwarding, first voting, and finalization.
The result shows an increase of at least a factor 5 for a cross-chain call, depending on the number of $\mathit{blocksPerPhase}$.
Given the current blockchain time of about 15 seconds per block, the transaction takes 45 minutes if $\mathit{waitingBlocks}$ is set to 30 and $\mathit{blocksPerPhase}$ is set to 10, compared to 7.5 minutes if invoked manually.

\begin{table}
	\centering
	\caption{Optimal/Average Time Consumption on Rinkeby in Blocks, Optimal/Average Time in Minutes (15 Seconds/Block)}
	\label{tbl:evaluation_time_optimal_rinkaby}
	\begin{tabular}{lr|rrrrrr}
		\multirow{6}{*}{\rotatebox[origin=c]{90}{blocksPerPhase}} & & \multicolumn{6}{c}{waitingBlocks} \\
		& & \multicolumn{2}{c}{\textbf{30}}  & \multicolumn{2}{c}{\textbf{50}} & \multicolumn{2}{c}{\textbf{100}} \\
		& & \textbf{blocks} & \textbf{min} & \textbf{blocks} & \textbf{min} & \textbf{blocks} & \textbf{min} \\
		\cline{2-8}
		& \textbf{5} & 164/168 & 41/42 & 264/268 & 66/67 & 514/518 & 129/130 \\ 
		& \textbf{10} & 174/180 & 44/45 & 274/278 & 69/70  & 524/528& 131/132 \\ 
		& \textbf{30} & 214/218 & 54/55 & 314/318 & 79/80 & 564/589 & 141/147 \\ 
		\cline{2-8}
		& \textbf{m} & \multicolumn{1}{c}{30} & 7.5 & \multicolumn{1}{c}{50} & 12.5 & \multicolumn{1}{c}{100}  & 25\\
	\end{tabular}
\end{table}

Considering the retrieved values given in Table~\ref{tbl:evaluation_time_optimal_rinkaby}, the transaction pool has a consistent and negligible impact on the query of one block, when enough gas is provided (i.e.,~at least the average paid per computational step in the previous blocks). 
Only for two calls, we encountered outliers due to network latency, one by $\mathit{blocksPerPhase}=10$ and $\mathit{waitingBlocks}=30$ with 184 instead of 178 and another by $\mathit{blocksPerPhase}=30$ and $\mathit{waitingBlocks}=100$ with 631 instead of 568.
Overall, the needed time is within an acceptable range for many cross-chain transactions, but may be critical for some transactions. Reducing the number of $\mathit{waitingBlocks}$ and $\mathit{blocksPerPhase}$ (with a trade-off regarding security) improves the time in situations where a faster communication is required.

\section{Conclusion}
\label{conclusion}
\noindent
In this paper, we presented a method for invoking smart contracts across blockchains.
The proposed method supports not only a single cross-chain smart contract call, but also the recursive invocation as well as the continuation of a smart contract after receiving a result.
Our solution takes into account the concurrency and consistency of blockchains and the payment of transaction fees and rewards.
The proposed implementation has been evaluated with several tests where results have shown a time overhead by a factor of at least 5 and a cost overhead by a factor of 40.
In future work, we will investigate ways to reduce the cost of cross-chain calls.
Possible options include the use of off-chain transactions, the optimization of source code, or a mixed approach with relay networks.
Furthermore, we want to improve the voting mechanism by increasing the incentives to vote correctly, add support for token transfers and support further blockchains.

\section*{Acknowledgment}
\noindent
The financial support by the Vienna Science and Technology Fund (WWTF) grant VRG18-013, the Austrian Federal Ministry for Digital and Economic Affairs, the National Foundation for Research, Technology and Development and the Christian Doppler Research Association is gratefully acknowledged.

\bibliographystyle{IEEEtran}
\bibliography{IEEEabrv,main}

\end{document}